\documentclass[prb,aps,tightenlines,twocolumn,superscriptaddress,
showpacs,floatfix]{revtex4}
\usepackage{colordvi}
\usepackage{graphicx}
\usepackage{amssymb}
\usepackage{amsmath}

\begin{document}
\title{Electron spin relaxation in $p$-type GaAs quantum wells} 
\author{Y. Zhou}
\affiliation{Hefei National Laboratory for Physical Sciences at
  Microscale, University of Science and Technology of China, Hefei,
  Anhui, 230026, China} 
\author{J. H. Jiang}
\affiliation{Department of Physics,
  University of Science and Technology of China, Hefei,
  Anhui, 230026, China}
\author{M. W. Wu}
\thanks{Author to whom correspondence should be addressed}
\email{mwwu@ustc.edu.cn.}
\affiliation{Hefei National Laboratory for Physical Sciences at
  Microscale, University of Science and Technology of China, Hefei,
  Anhui, 230026, China} 
\affiliation{Department of Physics,
  University of Science and Technology of China, Hefei,
  Anhui, 230026, China}

\date{\today}
\begin{abstract}
We investigate electron spin relaxation
in $p$-type GaAs quantum wells from a fully microscopic kinetic spin
Bloch equation approach, with all the relevant scatterings, such as
the electron-impurity, electron-phonon, electron-electron Coulomb,
electron-hole Coulomb and electron-hole exchange (the
Bir-Aronov-Pikus mechanism) scatterings  explicitly
included. Via this approach, we examine the relative importance of the
D'yakonov-Perel' and Bir-Aronov-Pikus mechanisms in wide ranges 
of temperature, hole density, excitation density and impurity
density, and present a phase-diagram--like picture showing the
parameter regime where the D'yakonov-Perel' or Bir-Aronov-Pikus
mechanism is more important. 
It is discovered that in the impurity-free case the temperature regime
where the Bir-Aronov-Pikus mechanism is more efficient than the
D'yakonov-Perel' one is around the hole Fermi temperature for
high hole density, regardless of excitation density.
However, in the high impurity density case with the impurity density
being identical to the hole density, this regime is roughly
from the electron Fermi temperature to the hole Fermi temperature. 
Moreover, we predict that for the impurity-free case, in
the regime where the D'yakonov-Perel' mechanism dominates the spin
relaxation at all temperatures, the temperature dependence of the spin
relaxation time presents a {\em peak} around the hole Fermi temperature,
which originates from the electron-hole Coulomb scattering.
We also predict that at low temperature, the hole-density dependence
of the electron spin relaxation time exhibits a {\em double-peak} structure in
the impurity-free case, whereas first a peak and then a valley in the case of
identical impurity and hole densities. These intriguing behaviors are due to 
the contribution from holes in high subbands.
\end{abstract}

\pacs{72.25.Rb, 67.30.hj, 71.10.-w}

\maketitle

\section{Introduction}
In recent years, much attention has been devoted to semiconductor
spintronics both theoretically and experimentally due to the potential
application of spin-based devices.\cite{opt-or,Wolf,spintronics}  
In order to manipulate the spin relaxation such that the
information is well preserved before required
operations are completed, it 
is crucial to gain a thorough understanding of spin relaxations. 
In $p$-doped III-V semiconductors, the main electron spin relaxation
mechanisms have been recognized as:\cite{opt-or} the
Bir-Aronov-Pikus (BAP) mechanism\cite{Bir} 
and the D'yakonov-Perel' (DP) mechanism.\cite{Dyakonov} 
In the DP mechanism, electron spins decay due to their
 precessions around the momentum-dependent spin-orbit fields 
(inhomogeneous broadening)\cite{Wu_01}
  during the free flight between adjacent scattering events.
In the BAP mechanism, spin relaxes due to spin-flip
  caused by exchange interaction with holes.
It was believed that in $p$-doped bulk samples the BAP mechanism
dominates the spin relaxation process at high doping density and low
temperature, whereas the DP mechanism is more important at low doping
density and high 
temperature.\cite{opt-or,Song,Aronov_83,Fishman,Zerrouati}
In two-dimensional system, Maialle\cite{Maialle_96}
calculated the spin relaxation time (SRT) due to these two mechanisms 
at zero temperature by using the single-particle approach and showed
that these two SRTs have nearly the same order of magnitude.  However,
as pointed out by Zhou and Wu lately,\cite{Zhou_PRB_08}
there are some common problems in the previous literature:
The SRT due to the BAP mechanism was calculated based on the 
elastic scattering approximation, which is invalid at 
low temperature due to the omission of the Pauli blocking. 
Also, the investigation of the SRT due to the DP mechanism was also
inadequate because the Coulomb scattering
is not included in the frame of the single-particle theory.

Zhou and Wu applied the fully microscopic kinetic spin Bloch equation
(KSBE) approach\cite{Wu_01,Wu_rev} to investigate the spin relaxation
in $p$-type GaAs quantum wells.\cite{Zhou_PRB_08}  
The KSBE approach has achieved good
success in the study of the spin dynamics in semiconductors, where 
not only the results are in good agreement with the previous experiments, 
but also many predictions have been confirmed by the latest experiments.  
\cite{Wu_01,Wu_rev,Zhou_PRB_08,wu-strain,highP,hot-e,Zhou_PRB_07,Yang_exp,
wu-exp,wu-exp-hP,multi-valley,Jiang_GaMnAs_09,nGaAs-peak,Jiang_bulk_09,
Lai,Awsch-exp,Ji,Zheng,tobias,peak-cpl}  
Via this approach, they explicitly included all the
relevant scatterings and obtained the accurate SRT due to these two mechanisms. 
It was found that the BAP mechanism is always less efficient than the DP
mechanism for moderate and high excitation densities
where $N_{\rm ex}\gtrsim 0.1 N_h$ [$N_{\rm ex}$ ($N_h$) is the
excitation (hole) density], in contrast to the common belief in the
previous literature.\cite{opt-or,Song,Aronov_83,Fishman,Zerrouati}
This claim has very recently been confirmed experimentally
by Yang {\em et al.}.\cite{Yang_exp} Moreover, a similar 
conclusion was also obtained in bulk GaAs later.\cite{Jiang_bulk_09} 

However, for very low excitation density where the Pauli blocking of
electrons is negligible, for high hole density where the 
contribution from the high subbands or different hole bands 
becomes significant and/or for high impurity density where the spin
relaxation due to the DP mechanism is 
suppressed, whether the BAP mechanism can be more efficient is
still questionable. In the present work, we extend  the KBSEs to
 include  both the lowest subband of
light-hole (LH) and the lowest two subbands of heavy-hole (HH),
and compare the relative importance of 
the DP and BAP mechanisms in wider ranges of temperature, hole 
density, excitation density and impurity density. 
We present a ``phase-diagram--like'' picture indicating the dominant spin
relaxation mechanism. 
In the case with no impurity and high excitation density, our 
results  show that the BAP mechanism is unimportant at low
temperature, in consistence with Ref.~\onlinecite{Zhou_PRB_08}.
Nevertheless, since more hole subbands and 
bands are included in our model, we are able to
discuss the case with higher hole density. We find that the BAP
mechanism can surpass the DP mechanism at {\em high} temperature for
{\em sufficiently } high hole density. 
In the case with no impurity and low excitation density, the BAP
mechanism can surpass the DP mechanism for wider hole-density and
temperature ranges. Moreover, we also find that in both cases above,
the regime where the BAP mechanism is more efficient  is always around
the hole Fermi temperature for high hole density, regardless of
excitation density. 
However, in the high impurity density case with the impurity density
being identical to the hole density, the behavior is very
different from the impurity-free case:  the regime of hole density
where the BAP mechanism is more efficient becomes larger, and
 the regime of temperature becomes wider,
ranging roughly from the electron
Fermi temperature to the hole Fermi temperature. 
We also show that the multi-hole-subband effect leads to
a very intriguing hole-density dependence of SRT at low
temperature.

This paper is organized as follows. In Sec.~II, we set up the KSBEs. 
In Sec.~III, we compare the relative importance of the BAP and DP
mechanisms in different parameter regimes and investigate the 
multi-hole-subband effect. We conclude in Sec.~IV. 
 A comparison of the calculation from the KSBEs
 with the experimental data of a $p$-type GaAs quantum well is
given in the Appendix.

\section{Kinetic Spin Bloch Equations}
We investigate a $p$-type (001) GaAs quantum well of width $a$ with its growth
direction along the $z$-axis. The width is assumed to be small
enough so that only the lowest subband of electron, the lowest
two subbands of HH and the lowest subband of LH
are relevant for the electron and hole densities we discuss. 
The envelope functions of the relevant subbands are calculated under
the finite-well-depth assumption.\cite{Zhou_PRB_07,Zhou_PRB_08} 
The barrier layer is chosen to be Al$_{0.4}$Ga$_{0.6}$As where the
barrier heights of electron and hole are 328 and 177~meV,
respectively.\cite{Yu_92} We focus on the metallic regime where most
of the carriers are in extended states. Since the hole spins relax
very rapidly (only several picoseconds), we assume that the hole
system is always in the equilibrium.

Via the nonequilibrium Green function method,\cite{Haug_1998} we
construct the KSBEs as follows:\cite{Wu_rev,Zhou_PRB_08}
\begin{equation}
  \partial_t\hat{\rho}_{{\bf k}}=
  \partial_t\hat{\rho}_{{\bf k}}|_{\rm coh}
  +\partial_t\hat{\rho}_{{\bf k}}|_{\rm scat},
\end{equation}
with $\hat{\rho}_{\bf k}$ representing the electron single-particle
density matrix with a two-dimensional momentum ${\bf k}=(k_x,k_y)$, 
whose diagonal and off-diagonal elements describe the
electron distribution function and spin coherence
respectively. The coherent term can be written as
($\hbar\equiv 1$ throughout this paper)
\begin{equation}
\partial_t\hat{\rho}_{{\bf k}}|_{\rm coh}=
-i\left[{\bf h}({\bf k})\cdot\frac{\hat{\mbox{\boldmath$\sigma$}}}{2}
+\hat{\Sigma}_\mathrm{HF}({\bf k}),\;\; \hat{\rho}_{{\bf k}} \right],
\end{equation}
in which $[A,B]\equiv AB-BA$ is the commutator. ${\bf h}({\bf k})$ represents the
spin-orbit coupling of electrons composed of the
Dresselhaus\cite{Dresselhaus_55} and Rashba\cite{Rashba_84} terms.
For GaAs quantum wells, the Dresselhaus term is dominant\cite{Lau_05} and 
\begin{eqnarray}
  {\bf h}({\bf k})=2\gamma_{\rm D}\Big(\,k_x(k_y^2-\langle k_z^2 \rangle ),\
  k_y(\langle k_z^2 \rangle-k_x^2),\ 0\,\Big),
\end{eqnarray}
where $\langle k_z^2 \rangle$ stands for the average of the operator
$-(\partial / \partial z)^2$ over the state of the lowest
subband of electron, and $\gamma_{\rm D}= 8.6$~eV$\cdot${\AA}$^3$ 
denotes the Dresselhaus spin-orbit coupling
coefficient.\cite{SOC1,wu-exp} 
$\hat{\Sigma}_\mathrm{HF}({\bf k})$ is the effective magnetic field
from the Coulomb Hartree-Fock contribution.\cite{highP} 
For the screened Coulomb potential,  the screening from
electrons and holes is calculated under
the random phase approximation.\cite{Zhou_PRB_08,mahan}    
The scattering term $\partial_t\hat{\rho}_{{\bf k}}|_{\rm scat}$
consists of the electron-impurity, electron-phonon, electron-electron
Coulomb, electron-hole Coulomb, and electron-hole exchange
scatterings. The expressions of these scatterings are given in detail
in Ref.~\onlinecite{Zhou_PRB_08}. Here we just extend the
electron-hole Coulomb and exchange scatterings to the
multi-hole-subband case. The expression of electron-hole Coulomb
scattering is still similar to that in Ref.~\onlinecite{Zhou_PRB_08}.  
The complete electron-hole exchange scattering term is written as 
\begin{eqnarray}
    \nonumber
&& \left.{\partial}_t\hat{\rho}_{{\bf k}}
    \right|_{\rm BAP} = -\pi \hspace{-0.2cm} \sum_{{\bf k}^{\prime}{\bf q}\lambda
      \lambda^{\prime} \atop \eta=\pm}\hspace{-0.2cm}
    \delta(\epsilon_{{\bf k}-{\bf q}}-\epsilon_{\bf k}
    +\epsilon_{{\bf k}^{\prime},\lambda}^h
    -\epsilon_{{\bf k}^{\prime}-{\bf q},\lambda^{\prime}}^h)
    \\ \nonumber 
&&\mbox{}\times |{\cal T}^\eta_{\lambda\lambda^\prime}({\bf k}
    +{\bf k}^{\prime}-{\bf q})|^2
    \left[\hat{s}_{\eta}\hat{\rho}_{{\bf k}-{\bf q}}^{>}
      \hat{s}_{-\eta}\hat{\rho}_{{\bf k}}^{<}
    \right. 
(1-f^h_{{\bf k}^{\prime},\lambda})f^h_{{\bf k}^{\prime}-{\bf q},\lambda^{\prime}}\nonumber\\
&&\mbox{} - \hat{s}_{\eta}\hat{\rho}_{{\bf k}-{\bf q}}^{<}
      \hat{s}_{-\eta}\hat{\rho}_{{\bf k}}^{>}
 f^h_{{\bf k}^{\prime},\lambda}(1-f^h_{{\bf k}^{\prime}-{\bf q},\lambda^{\prime}})
    \Big]  + {\rm h.c.}\ .
  \label{equ:rho_bap}
\end{eqnarray}
Here $\hat{\rho}^{>}_{\bf k}=1-\hat{\rho}_{\bf k}$ and
$\hat{\rho}^{<}_{\bf k}=\hat{\rho}_{\bf k}$ are the electron density
matrices; $\hat{s}_{\pm}=\hat{s}_x\pm i\hat{s}_y$ are the electron spin
ladder operators. 
$\lambda={\rm HH}^{(n)},{\rm LH}^{(n)}$ with the superscript
being the subband index of hole. $f^h_{{\bf k},\lambda}$ is the hole
distribution on the $\lambda$th hole band.
The matrix $\hat{{\cal T}}^{\pm}$ comes from the long-range term of
the electron-hole exchange interaction Hamiltonian and can be written as
$\hat{{\cal T}}^{\pm}=\frac{3}{8}\frac{\Delta E_{\rm LT}} {|\phi_{3D}(0)|^2}
\hat{M}^{\pm}$,\cite{Maialle_93,short-range}
where $\Delta E_{\rm LT}$ is the longitudinal-transverse
splitting in bulk; $|\phi_{3D}(0)|^2=1/(\pi a_0^3)$ with $a_0$ being
the exciton Bohr radius; $\hat{M}^{-}$ and $\hat{M}^{+}$
($=(\hat{M}^{-})^{\dag}$) are operators in hole spin space.  
The matrix $\hat{M}^{-}$ is given by\cite{Maialle_93}  
(in the order of $|\frac{3}{2}\rangle^{(1)}$,
$|-\frac{3}{2}\rangle^{(1)}$, $|\frac{3}{2}\rangle^{(2)}$,
$|-\frac{3}{2}\rangle^{(2)}$, $|\frac{1}{2}\rangle^{(1)}$,
$|-\frac{1}{2}\rangle^{(1)}$)
\begin{equation}
\setlength{\arraycolsep}{0.6mm}
\hat{M}^{-}({\bf K})=\hspace{-0.1cm}\left[\hspace{-0.03cm}
\begin{array}{cccccc}
    0 & 0 & 0 & 0 & 0 & 0 \\
    F^{0}_{\mathtt{h1h1}}K_{+}^2 & 0 & 0 & 0 & 0 & 
    \frac{-F^{0}_{\mathtt{h1l1}}}{\sqrt{3}}K^2\\
    0 & 0 & 0 & 0 & 0 & 0 \\
    0 & 0 & F^{0}_{\mathtt{h2h2}}K_{+}^2 & 0 & 
    \frac{-2F^{1}_{\mathtt{h2l1}}}{\sqrt{3}}K_{+} & 0 \\
    \frac{-F^{0}_{\mathtt{l1h1}}}{\sqrt{3}}K^2 & 0 & 0 & 0 & 0 & 
    \frac{F^{0}_{\mathtt{l1l1}}}{3}K_{-}^2\\
    0 & 0 & \frac{2F^{1}_{\mathtt{l1h2}}}{\sqrt{3}}K_{+} & 0 & 
    \frac{-4F^{2}_{\mathtt{l1l1}}}{3} & 0 
\end{array}
\hspace{-0.03cm}\right]
\label{M_bap}
\end{equation}
where ${\bf K}={\bf k}+{\bf k}^\prime-{\bf q}$ is the center-of-mass momentum
of the electron-hole pair with $K_{\pm}=K_x \pm iK_y$. The form factors can be written as
\begin{equation}
F_{\lambda\lambda^\prime}^{p}(K)=\int\;\frac{dq_z}{2\pi}
\frac{q_z^p}{K^2+q_z^2} f_{\lambda\lambda^\prime}(q_z)
\end{equation}
with 
\begin{equation}
  f_{\lambda\lambda^\prime}(q_z)=\int\;dzdz^\prime\;\xi_e(z^\prime)\zeta_h^{\lambda^\prime}
  (z^\prime) e^{iq_z(z-z^\prime)} \zeta_h^\lambda(z)\xi_e(z).
\end{equation}
In Eq.~(\ref{M_bap}), it is seen that most of the nonzero elements in matrix
$\hat{M}^-$ contain $K_{\pm}^2$ or $K_{\pm}$, and thus the magnitudes
of these terms increase with increasing $K$. The only exception
is $M^-_{-\frac{1}{2},\frac{1}{2}}$, where the $K$ dependence 
is only from the form factor $F^{0}_{\mathtt{l1l1}}$.
Consequently the magnitude of $M^-_{-\frac{1}{2},\frac{1}{2}}$
decreases with $K$. This $K$ dependence
contributes to an intriguing hole density dependence of
spin relaxation to be addressed in this work.

\begin{table}[ht]
\caption{Material parameters used in the calculation}
\begin{center}
\begin{tabular}{llll}
  \hline \hline
  $g_e$ & $-0.44$ & $m_e^{\ast}$ & $0.067m_0$ \\
  $m_{{\rm HH},\parallel}^\ast$ & $0.112m_0$ & 
  $m_{{\rm HH},\perp}^\ast$ & $0.377m_0$ \\
  $m_{{\rm LH},\parallel}^\ast$ & $0.211m_0$ & 
  $m_{{\rm LH},\perp}^\ast$ & $0.091m_0$ \\
  $\kappa_0$ & 12.9 &  $\kappa_{\infty}$ & 10.8\\
  $D$ & $5.31\times 10^3$~kg/m$^3$ & $e_{14}$ & $1.41\times 10^9$~V/m\\
  $v_{st}$ & $2.48\times 10^3$~m/s & $v_{sl}$ & $5.29\times 10^3$~m/s \\
  $\Xi$ & 8.5~eV & $\omega_{\rm LO}$ &  35.4~meV \\
  $\Delta_{\rm SO}$ & 0.341~eV & $E_g$ &  1.55~eV \\
  $\Delta E_{\rm LT}$ & $0.08$~meV & $a_0$ & $146.1$~\AA \\
  \hline \hline
\end{tabular}
\end{center}
\label{parameter}
\end{table}

\section{Results and Discussions}

By numerically solving the KSBEs with all the scatterings explicitly
included, one is able to obtain the SRT from the temporal evolution
of the electron spin polarization along the $z$-axis.
We choose initial spin polarization $P=4$~\%\cite{polarize} and
well width $a=10$~nm, external magnetic field $B=0$ unless otherwise
specified. The other material parameters are listed in
Table~\ref{parameter}.\cite{para,Ekardt,Maialle_96}

\begin{figure}[tbp]
  \begin{center}
    \includegraphics[width=8.6cm]{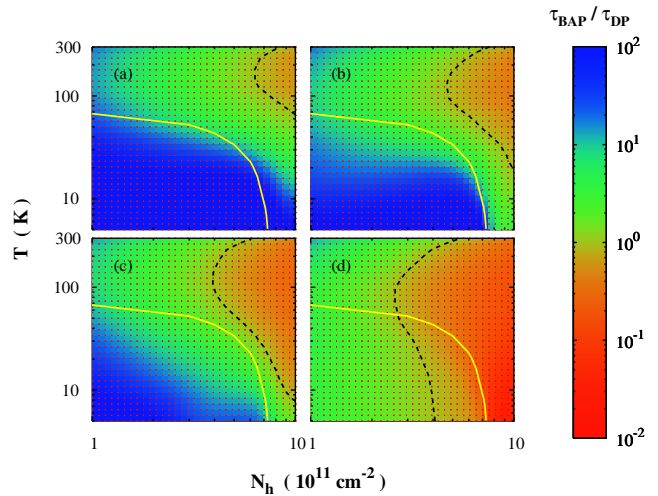}
  \end{center}
   \caption{(Color online) Ratio of the SRT due to the BAP mechanism
     to that due to the DP mechanism, $\tau_{\rm BAP}/\tau_{\rm DP}$,
     as function of temperature and hole density with 
     (a) $N_i=0$, $N_{\rm ex}=10^{11}$~cm$^{-2}$;
     (b) $N_i=0$, $N_{\rm ex}=10^{9}$~cm$^{-2}$; (c) $N_i=N_h$, 
     $N_{\rm ex}=10^{11}$~cm$^{-2}$; (d) $N_i=N_h$, $N_{\rm ex}=10^{9}$~cm$^{-2}$.
     The black dashed curves indicate the cases satisfying 
     $\tau_{\rm BAP}/\tau_{\rm DP}=1$. Note the smaller the
     ratio $\tau_{\rm BAP}/\tau_{\rm DP}$ is, the more
     important the BAP mechanism becomes.
     The yellow solid curves indicate the cases satisfying 
     $\partial_{\mu_h}[N_{{\rm LH}^{(1)}}+N_{{\rm HH}^{(2)}}]/\partial_{\mu_h}N_h=0.1$.
     In the regime above the yellow curve the multi-hole-subband
     effect becomes significant.
   }
  \label{fig_phase}
\end{figure}

\begin{figure}[tbp]
  \begin{center}
    \includegraphics[width=8.5cm]{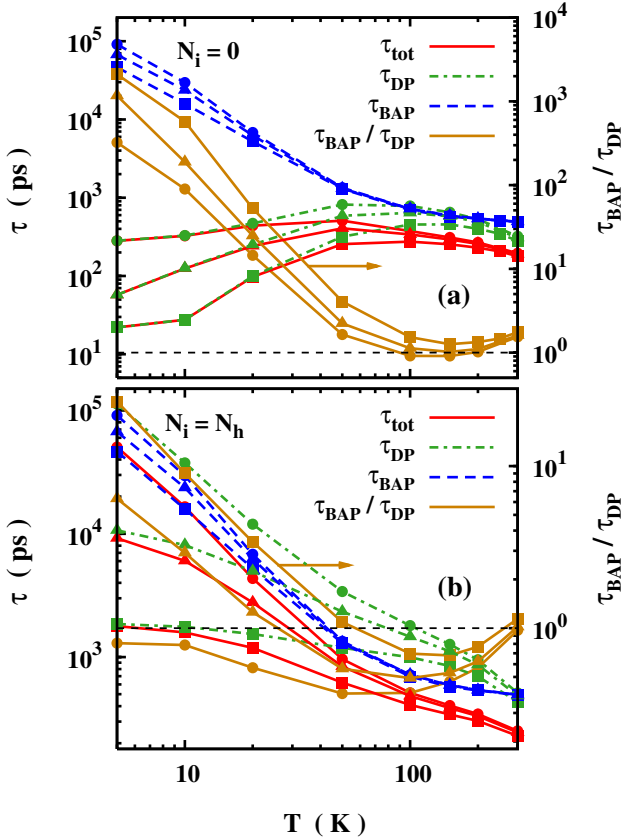}
  \end{center}
  \caption{(Color online)  SRTs due to the DP and BAP mechanisms, the
    total SRT together with the ratio $\tau_{\rm BAP}/\tau_{\rm DP}$
    {\em vs.} temperature $T$ for $N_{\rm ex}=10^{9}$~cm$^{-2}$ (curves with
    $\bullet$), $3\times 10^{10}$~cm$^{-2}$ (curves with
    $\blacktriangle$) and $10^{11}$~cm$^{-2}$ (curves with
    $\blacksquare$) with hole density $N_{h}=5 \times 10^{11}$~cm$^{-2}$ and
    impurity densities (a) $N_i=0$ and (b) $N_i=N_h$.
    The electron Fermi temperatures for those excitation
    densities are $T_{\rm F}^e=0.41$, $12.4$ and $41.5$~K, respectively. The
    hole Fermi temperature is $T_{\rm F}^h=124$~K. Note the scale of
    $\tau_{\rm BAP}/\tau_{\rm DP}$ is on the right-hand side
    of the frame. The multi-hole-subband effect is taken into
      account in the calculation.
  }
  \label{fig_Ne}
\end{figure}

\subsection{Comparison of the BAP and DP mechanisms}
We first examine the relative importance of the BAP and DP mechanisms
for different parameters in $p$-type GaAs quantum wells.
In Fig.~\ref{fig_phase}, the ratio of the SRT due to the BAP mechanism
to that due to the DP mechanism is plotted as function of
temperature and hole density in the cases with no/high impurity 
and low/high excitation densities. 
From this figure, one can recognize the parameter regime where the DP
or BAP mechanism is more important.
It is also shown that the multi-hole-subband effect becomes
significant for high temperature and/or high hole density (the regime
above the yellow solid curve). 
Here and hereafter, the multi-hole-subband  refers to either the
high HH subband or the LH subband. 
Although the multi-hole-subband effect has important effect on
electron spin relaxation in the relevant regime, the main physics is
still the same as that in the single-hole-subband model. Therefore,
in this subsection, we first discuss the general behavior about how 
the relative importance of the BAP and DP mechanisms is influenced
by the temperature, hole density, excitation density and impurity
density, which is analogous in both the multi-hole-subband and
single-hole-subband models. We then investigate the special features
from the contribution of high hole subbands in next subsection. 

In the case with no impurity and high excitation density 
[Fig.~\ref{fig_phase}(a)], our results are consistent with
Ref.~\onlinecite{Zhou_PRB_08}: i.e., the BAP mechanism is
unimportant at low temperature, which is in stark contrast with the
common belief in the literature.\cite{opt-or,Song,Aronov_83,Fishman,Zerrouati}
Moreover, since we extend the scope of our
  investigation to higher hole density by 
including more hole subbands in our model, 
it is discovered that the BAP mechanism can surpass the DP
mechanism in the regime with high temperature and sufficiently high
hole density (the regime embraced by the black dashed curve). 

In the case with no impurity and low excitation
density [Fig.~\ref{fig_phase}(b)], one can see 
that the regime where the BAP mechanism surpasses the DP mechanism
becomes larger. The underlying physics is shown in Fig.~\ref{fig_Ne}(a).
It is seen that the SRTs due to the BAP and DP mechanisms both decrease with 
increasing excitation density ($N_{\rm ex}$=$N_{e}$),
but the amplitude of the latter is much larger than the former.
The decrease of $\tau_{\rm DP}$ comes from the increase of the
inhomogeneous broadening $\langle |{\bf h}_{\bf k}|^2 \rangle\propto 
N_{\rm ex}$,\cite{strong_scat,highP} and the decrease of 
$\tau_{\rm BAP}$ is mainly from the increase of the average electron
velocity $\langle v_k \rangle\propto N_{\rm ex}^{0.5}$.\cite{Maialle_96} 
Moreover, the increase of the Pauli blocking of 
electrons can partially compensate the
effect of the increase of $\langle v_k \rangle$.\cite{Zhou_PRB_08} 
Consequently, $\tau_{\rm BAP}$ decreases with $N_{\rm ex}$ much more
slowly than $\tau_{\rm DP}$ and the relative
importance of $\tau_{\rm BAP}$ is enhanced for lower excitation density.
It is also noted that when the electron system is in the nondegenerate
regime ($T>T_{\rm F}^{e}=E_{\rm F}^{e}/k_{\rm B}$), the inhomogeneous
broadening and $\langle v_k \rangle$ is not sensitive to 
$N_{\rm ex}$. Thus the ratio $\tau_{\rm BAP}/\tau_{\rm DP}$ changes
little with the excitation density.

By comparing Fig.~\ref{fig_phase}(a) and (b), it is seen that the regimes
where the BAP mechanism is more efficient in both cases are
around the hole Fermi temperature $T_{\rm F}^h=E_{\rm F}^{h}/k_{\rm B}$
for high hole density.  Here $E_{\rm F}^{h}$ represents the Fermi
energy of hole at zero temperature calculated with the HH$^{(1)}$,
LH$^{(1)}$ and HH$^{(2)}$ subbands included. A typical case is shown in
Fig.~\ref{fig_Ne}(a) for $N_h=5\times 10^{11}$~cm$^{-2}$.\cite{Nh}   
It is shown that the ratio $\tau_{\rm BAP}/\tau_{\rm DP}$ first
decreases and then increases with increasing $T$.\cite{weak_scat}
The minimum is around $T_{\rm F}^h=124$~K, regardless of excitation density.
The underlying physics is as follows. On one hand, the
SRT due to the DP mechanism first increases and then decreases with
$T$ and the peak appears around the hole Fermi temperature.   
This is because the electron-hole Coulomb scattering, which dominates
the momentum scattering, increases with increasing temperature in  the degenerate
regime ($T<T_{\rm F}^{h}$) and decreases with $T$ in the nondegenerate
regime ($T>T_{\rm F}^{h}$), similar to the electron-electron Coulomb
scattering.\cite{Zhou_PRB_07,Ivchenko,Vignale}
On the other hand, the SRT due to the BAP mechanism first decreases
rapidly and then slowly with $T$. The decrease of $\tau_{\mathrm{BAP}}$
is mainly from the decrease of the Pauli blocking of holes and the
increase of the matrix elements in Eq.~(\ref{M_bap}).\cite{Zhou_PRB_08} 
In high temperature (nondegenerate) regime, the Pauli blocking becomes
very weak, and thus $\tau_{\mathrm{BAP}}$ decreases slowly with $T$. 
Under the combined effect of these two mechanisms, the valley in the
ratio $\tau_{\rm BAP}/\tau_{\rm DP}$ appears around $T_{\rm F}^h$.   

Moreover, we also show that in the regime where the DP mechanism is
dominant at all temperatures, e.g., the high excitation
density case [the curves with squares in Fig.~\ref{fig_Ne}(a)], the
 total SRT shows a peak around the hole Fermi 
temperature. This temperature dependence is similar to the peak first
predicted theoretically and then confirmed experimentally in 
$n$-type samples.\cite{Franz,Zhou_PRB_07,Ji}   
The only difference is that the peak in the previous work comes from the
electron-electron Coulomb scattering and hence appears around the 
electron Fermi temperature, whereas the peak here originates
from the electron-hole Coulomb scattering and thus appears around the
hole Fermi temperature. 

Then we turn to the case of high impurity density with $N_i=N_h$
[Fig.~\ref{fig_phase}(c) and (d)]. 
In this case, the regime where the BAP mechanism
is more important becomes  
larger than that in the impurity-free case. The scenario is that
the higher impurity density strengthens the electron-impurity
scattering and suppresses the DP mechanism, consequently enhances the
relative importance of the BAP mechanism. 
Interestingly, it is also seen that the temperature regime where the BAP
mechanism surpasses the DP mechanism in this case is very different
from that in the impurity-free case. This regime
is roughly from the electron Fermi temperature to the hole Fermi 
temperature for high hole density.
To explore the underlying physics, we plot the SRTs due to the BAP and
DP mechanisms in Fig.~\ref{fig_Ne}(b) for $N_h=5\times 10^{11}$~cm$^{-2}$.
It is seen that the SRT due to the DP mechanism first decreases slowly
and then rapidly with temperature. This is because the
electron-impurity scattering, which dominates the momentum scattering,
has a very weak temperature dependence. Thus the temperature
dependence of $\tau_{\rm DP}$ is mainly determined by the
inhomogeneous broadening from the spin-orbit coupling. 
It is also noted that the inhomogeneous broadening changes little with
temperature when $T<T_{\rm F}^e$, hence $\tau_{\rm DP}$ varies with
$T$ very mildly at low temperature. 
On the contrary, as mentioned above, the SRT due to the BAP
mechanism first decreases rapidly and then slowly with temperature. 
As a result, the temperature dependence of $\tau_{\mathrm{BAP}}/\tau_{\mathrm{DP}}$
can be easily understood. When $T<T_{\rm F}^e$, $\tau_{\mathrm{DP}}$ decreases
with $T$ slower than $\tau_{\mathrm{BAP}}$, thus the ratio decreases
with $T$. In the case with $T>T_{\rm F}^h$, $\tau_{\mathrm{DP}}$ decreases
with $T$ faster than $\tau_{\mathrm{BAP}}$, hence the ratio
increases with $T$. The ratio $\tau_{\rm BAP}/\tau_{\rm DP}$ varies
mildly when temperature varies from $T_{\rm F}^e$ to $T_{\rm F}^h$. 
Consequently, when hole density is high enough, the BAP mechanism
can surpass the DP mechanism in the temperature regime between these two
temperatures. In particular, in the case with high impurity and very
low electron excitation densities [Fig.~\ref{fig_phase}(d)], the
electron Fermi temperature (0.41~K) is much lower than the
lowest temperature (5~K) of our computation and the hole Fermi
temperature is close to the highest temperature (300~K) of our computation. 
As a result, the BAP mechanism dominates the spin relaxation in the
{\em whole temperature regime} of our investigation. 

We stress that the different behaviors in the impurity-free and
high impurity density cases originate from the {\em different dominant
momentum scatterings}: the electron-hole Coulomb scattering in the
impurity-free case and the electron-impurity scattering in the
high impurity density case. The different dominant
scatterings  lead to the different temperature dependences of  
$\tau_{\rm DP}$, and hence the different 
behaviors of the ratio $\tau_{\rm BAP}/\tau_{\rm DP}$. 
In the case with moderate impurity density, these two scatterings both
contribute to the DP spin relaxation, thus the trend of the
temperature dependence of $\tau_{\rm DP}$ is  between
those in the impurity-free and high impurity density cases. As a
result, the temperature regime where the BAP mechanism is more
efficient than the DP mechanism is from some temperature between the
electron and hole Fermi temperatures to the hole Fermi temperature.

\subsection{Multi-hole-subband effect}
Now we investigate the multi-hole-subband effect on the spin
relaxation. In our model, besides the first HH subband, we also
consider the contribution from the first LH subband
and the second HH subband. 
Since only the hole states around the Fermi surface 
 can contribute to the
electron-hole Coulomb or exchange scattering, we choose
$\partial_{\mu_h}N_\lambda/\partial_{\mu_h}N_h$ as the criterion of
the contribution from $\lambda$ hole subband.  
We further show the regime where the contribution from high hole
subbands becomes significant in Fig.~\ref{fig_phase} (the regime above
the yellow curve), where $\partial_{\mu_h}(N_{{\rm LH}^{(1)}}+
N_{{\rm HH}^{(2)}})/\partial_{\mu_h}N_h>0.1$. 
It is noted that we only discuss the combined effect 
of the DP spin relaxation from the LH$^{(1)}$ and HH$^{(2)}$
subbands in the following, as the effects on
the electron-hole Coulomb scattering from these two subbands are
analogous. Moreover, the matrix elements in Eq.~(\ref{M_bap}) relevant
to the HH$^{(2)}$ subband are one order of magnitude smaller than
those relevant to the LH$^{(1)}$ subband for the
relevant range of center-of-mass momentum $K$ in the following
cases. Therefore, we only discuss the 
effect on the BAP spin relaxation from the LH$^{(1)}$ subband. 

\begin{figure}[tbp]
  \begin{center}
    \includegraphics[width=8.5cm]{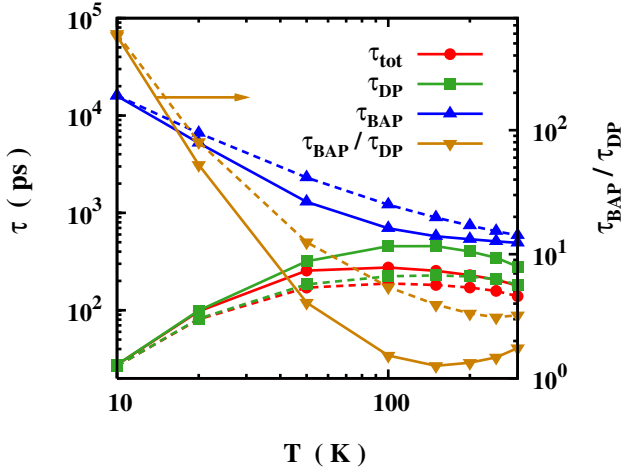}
  \end{center}
  \caption{(Color online) SRTs due to the DP and BAP mechanisms, the
    total SRT together with the ratio $\tau_{\rm BAP}/\tau_{\rm DP}$ {\em vs}.
    temperature $T$ with $N_i=0$, $N_h=5 \times 10^{11}$~cm$^{-2}$ and 
    $N_{\rm ex}=10^{11}$~cm$^{-2}$. The hole and electron Fermi temperatures are
    $T_{\rm F}^h=124$~K and $T_{\rm F}^e=41$~K, respectively. 
    The sold curves are the results calculated with the lowest two
    subbands of HH and the lowest subband of LH. The dash curves are
    those from only the lowest subband of HH. Note the scale of 
    $\tau_{\rm BAP}/\tau_{\rm DP}$ is on the right-hand side of the
    frame.
  }
  \label{fig_T}
\end{figure}

We first show how the multi-hole-subband effect influences the
temperature dependence of the spin relaxation. The SRTs due to the DP
and BAP mechanisms as well as the ratio $\tau_{\rm BAP}/\tau_{\rm DP}$ 
are plotted in Fig.~\ref{fig_T} as function of temperature for a
typical case with $N_i=0$, $N_h=5 \times 10^{11}$~cm$^{-2}$ and 
$N_{\rm ex}=10^{11}$~cm$^{-2}$. 
It is seen that after considering the contribution from high hole
subbands, $\tau_{\rm BAP}$ decreases but $\tau_{\rm DP}$ increases,
and hence the importance of the BAP mechanism is enhanced.  
The underlying physics is as follows. The states in high hole subbands
provide additional scattering channel, and the electron-hole Coulomb
and exchange scatterings are both enhanced. The former
suppresses the DP mechanism in the strong scattering limit, and the
latter leads to an enhancement of the BAP mechanism. Both make the BAP
mechanism become more important compared with the DP mechanism. 
It is also seen that the multi-hole-subband effect becomes more
pronounced  at high temperature. This is because the occupation of the
high hole subbands becomes larger when temperature increases. 

From Fig.~\ref{fig_T}, one also finds that the multi-hole-subband
effect does not significantly affect the trend of the temperature
dependence of the SRT. The main change after the inclusion of the
high hole subbands is that the temperature at which 
$\tau_{\rm BAP}/\tau_{\rm DP}$ reaches minimum becomes closer to the
hole Fermi temperature. 
The underlying physics is as follows.
In the degenerate regime ($T<T_{\rm F}^{h}$), it is seen that compared
with those in the single-hole-subband
model, $\tau_{\rm DP}$  ($\tau_{\rm BAP}$) in the multi-hole-subband model 
increases (decreases) faster with increasing temperature,
both originate from the increase in the occupation of
the high hole subbands and hence the increase of the electron-hole 
Coulomb and exchange scatterings.  This leads to a
faster decrease of $\tau_{\rm BAP}/\tau_{\rm DP}$ with increasing temperature
when $T<T_{\rm F}^h$.\cite{Jiang_GaMnAs_09,Jiang_bulk_09}
Nevertheless, in the nondegenerate regime ($T>T_{\rm F}^{h}$), it is seen that
$\tau_{\rm DP}$  ($\tau_{\rm BAP}$) 
in the multi-hole-subband model decreases faster (slower) than
that in the single-hole-subband model. The accelerating in the
decrease of $\tau_{\rm DP}$ can be understood as follows. 
In the nondegenerate regime, the
electron-hole Coulomb scattering decreases with temperature.
With the contribution from high hole subbands,
the electron-hole Coulomb scattering becomes stronger 
and thus the decrease rate also becomes larger. Therefore,
$\tau_{\rm DP}$ decreases faster in the multi-hole-subband
calculation.\cite{Jiang_GaMnAs_09,Jiang_bulk_09}
The slowdown in the decrease of $\tau_{\rm BAP}$ originates from the
anomalous $K$ dependence of the matrix element $M^-_{-\frac{1}{2},\frac{1}{2}}$ in
Eq.~(\ref{M_bap}), which is relevant to the LH$^{(1)}$ subband.
As discussed above, in the nondegenerate regime, the temperature dependence of 
$\tau_{\rm BAP}$ is mainly from the matrix elements.  
It is also noted that the magnitude of
$M^-_{-\frac{1}{2},\frac{1}{2}}$ decreases with $K$, whereas the 
magnitudes of the other matrix elements increase with $K$. With
the increase of temperature, more holes and electrons are distributed at larger
momentums, the contribution from $M^-_{-\frac{1}{2},\frac{1}{2}}$ decreases,
while the contributions from the other matrix elements increases. 
These two trends counteract with each other and make
$\tau_{\rm BAP}$ decrease with increasing $T$ very slowly at high temperature. 
Consequently, when $T>T_{\rm F}^h$, the ratio $\tau_{\rm BAP}/\tau_{\rm DP}$
shows a steeper increase with the rising temperature 
in the multi-hole-subband calculation.
Therefore, both trends when the temperature is below and above $T_{\rm F}^h$
make the minimum of  $\tau_{\rm BAP}/\tau_{\rm DP}$ appear at the 
temperature closer to $T_{\rm F}^h$ in the multi-hole-subband calculation. 

\begin{figure}[tbp]
  \begin{center}
    \includegraphics[width=8.5cm]{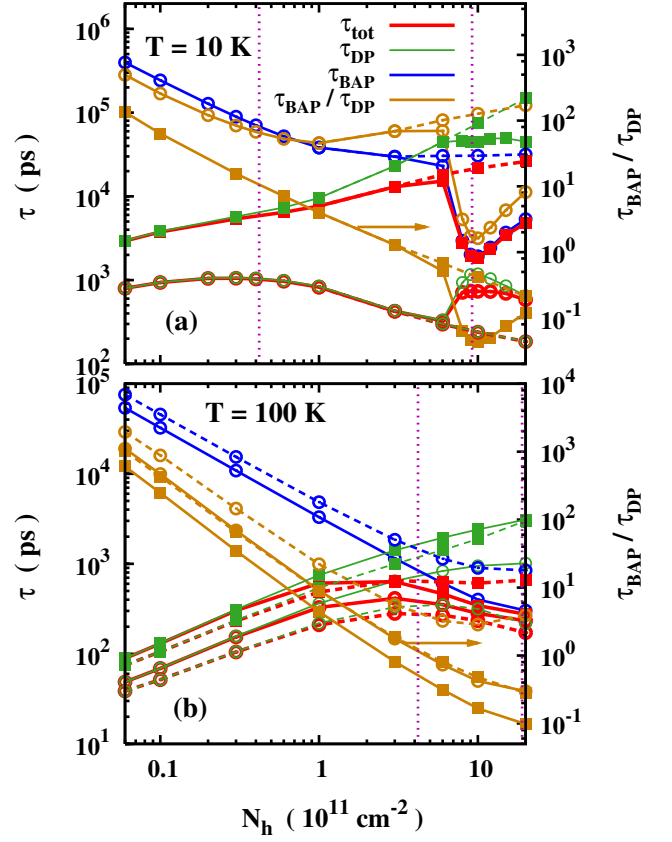}
  \end{center}
  \caption{(Color online) SRTs due to the DP and BAP mechanisms, the total
    SRT together with the ratio $\tau_{\rm BAP}/\tau_{\rm DP}$
   {\em vs.} hole density $N_h$ with impurity 
    densities $N_i=0$ ($\circ$) and $N_i=N_h$ ($\blacksquare$) at (a)
    $T=10$~K and (b) $100$~K (note that the scale of $\tau_{\rm
      BAP}/\tau_{\rm DP}$ is on the right-hand side of the frame). 
The excitation density $N_{\rm ex}=10^{9}$~cm$^{-2}$.
    The sold curves are the results
    calculated with the lowest two subbands of HH and the lowest subband
    of LH. The dashed curves are those calculated with only the lowest
    subband of HH. The two purple dotted vertical lines indicate the
   hole densities satisfying $E_{\rm F}^h=k_{\rm B}T$ and $E_{\rm F}^h-\Delta 
    E_{{\rm LH}^{(1)}}=k_{\rm B}T$, respectively [note that the second
    dotted line in (b) is very close to the  right frame].
    Here $\Delta E_{{\rm LH}^{(1)}}$ represents the energy splitting
    between the LH$^{(1)}$ and HH$^{(1)}$ subbands. 
  }
  \label{fig_Nh}
\end{figure}

We also investigate the multi-hole-subband effect on the hole-density
dependence of the spin relaxation. In Fig.~\ref{fig_Nh},
the SRTs due to various mechanisms, the total SRT together
with the ratio $\tau_{\rm BAP}/\tau_{\rm DP}$ are plotted as function of hole
density. It is interesting to see from Fig.~\ref{fig_Nh}(a)
that at low temperature,  the spin
relaxation has a very intriguing hole-density dependence. In the
impurity-free case, the hole-density dependence of the 
total SRT shows a double-peak structure: i.e., it  first increases
slightly, then decreases, again increases rapidly and finally decreases with 
increasing hole density. In the high impurity density case 
with $N_i=N_h$, the total SRT shows first a peak and then a
valley as a function of hole density.  
We first discuss the impurity-free case where the BAP mechanism is
negligible and the double-peak structure is solely from the DP mechanism.
The first peak can be understood as follows.
The electron-hole Coulomb scattering
increases with $N_h$ in the nondegenerate regime ($E_{\rm F}^h< 
k_{\rm B}T$) from the increase of the hole distribution, but 
decreases with $N_h$ in the degenerate regime ($E_{\rm F}^h>k_{\rm B}T$)
due to the increase of the Pauli 
blocking of holes.\cite{Jiang_GaMnAs_09,Jiang_bulk_09} 
Hence $\tau_{\rm DP}$ first increases and then decreases with
$N_h$ with the peak appearing around the hole density satisfying 
$E_{\rm F}^h=k_{\rm B}T$. This behavior is similar to the peak
predicted in $n$-type samples,\cite{Jiang_bulk_09} where the peak
originates from the electron-electron Coulomb scattering and hence 
appears around the electron density satisfying $E_{\rm F}^e=k_{\rm  B}T$. 
It is also seen that the second peak only appears in the
multi-hole-subband calculation, but becomes absent in the single-hole-subband
calculation (the green dashed curve with circles), which indicates that this peak
comes from the contribution of the electron-hole Coulomb scattering
from high hole subbands. 
In fact, the scenario is similar to the first one.
When $N_h>6\times10^{11}$~cm$^{-2}$, the contribution from the LH$^{(1)}$
subband becomes important.\cite{highHH} Since the holes in the
LH$^{(1)}$ subband are still in the nondegenerate regime, the
electron-hole Coulomb scattering increases with increasing 
hole density. Thus $\tau_{\rm DP}$ increases with $N_h$. 
When $N_h>9\times10^{11}$~cm$^{-2}$, i.e., 
$E_{\rm F}^h-\Delta E_{{\rm LH}^{(1)}}>k_{\rm B}T$ with
$\Delta E_{{\rm LH}^{(1)}}$ representing the energy splitting between the
LH$^{(1)}$ and HH$^{(1)}$ subbands, the holes in the 
LH$^{(1)}$ subband are also in the degenerate regime, and thus the effect of the
Pauli blocking becomes significant. Consequently $\tau_{\rm DP}$
decreases with $N_h$. The second peak appears around the
hole density satisfying $E_{\rm F}^h-\Delta E_{{\rm LH}^{(1)}}=k_{\rm
  B}T$. 

Then we turn to the case of high impurity density with the
impurity density being identical to the hole density. 
The scenario of the peak is as follows. The SRT due to the DP
mechanism increases monotonically with hole density, since the
electron-impurity scattering increases with $N_h$($=N_i$). 
Moreover, the SRT due to the BAP mechanism decreases with
$N_h$ due to the increase of the hole distribution of the HH$^{(1)}$
and LH$^{(1)}$ subbands, similar to the electron-hole Coulomb
scattering. As a result, the peak appears around the hole density 
where the BAP mechanism surpasses the DP
mechanism.\cite{Jiang_GaMnAs_09} 
It is also seen that $\tau_{\rm tot}$ increases with hole density when 
$N_h>9\times10^{11}$~cm$^{-2}$. The underlying physics is as follows. 
Similar to the previous discussion of the temperature
dependence, with the increase of hole density, the decrease of the
matrix element $M^-_{-\frac{1}{2},\frac{1}{2}}$ counteracts 
the increase of the other matrix elements. 
Thus the dependence of the hole density from the matrix elements is weak.
Consequently, when the holes in the HH$^{(1)}$ and LH$^{(1)}$ subbands
are both in the degenerate regime, the effect of the increase of the Pauli
blocking is dominant, and $\tau_{\rm BAP}$ increases with $N_h$.
The valley appears around the hole density satisfying 
$E_{\rm F}^h-\Delta E_{{\rm LH}^{(1)}}=k_{\rm B}T$. 
It is also noted that the increase of $\tau_{\rm BAP}$ only appears in
the multi-hole-subband calculation. In the single-hole-subband
calculation, $\tau_{\rm BAP}$ does not increase with $N_h$ but remains
almost a constant for high hole density (the blue dashed curve), since 
the increase of the Pauli blocking is counteracted by the
increase of the matrix elements relevant to the HH$^{(1)}$ subband. 

The hole-density dependence of the spin relaxation
at high temperature is also investigated [Fig.~\ref{fig_Nh}(b)].
Differing from the behavior at low temperature, it is seen that there 
is only one peak in both the impurity-free and high impurity density cases.  
We further show that the peaks in both cases come from the competition
of the DP and BAP mechanisms, which is similar to the peak in the case
with high impurity density and low temperature. 
The absence of the peak from the electron-hole Coulomb scattering 
is  due to the multi-hole-subband effect. As discussed above, when
the hole density is high enough so that the holes in the lowest
subband are in the degenerate regime, the contribution of the
electron-hole Coulomb scattering from the lowest hole subband
decreases with increasing hole density due to the increase of the Pauli blocking.
However, at high temperature the contribution from high hole subbands
is also important in this hole density regime. It is further noted that the
holes in the high subbands are still in the nondegenerate regime, 
thus the contribution from the high hole subbands increases rapidly
with $N_h$ and surpasses the effect from the
lowest hole subband. 
Consequently, $\tau_{\rm DP}$ continues to increase with $N_h$ and the
Coulomb peak disappears.

\section{Conclusion and Discussion}

In conclusion, we have performed a comprehensive investigation of 
electron spin relaxation in $p$-type GaAs quantum wells from a fully
microscopic KSBE approach. All relevant scatterings, such as, the
electron-impurity, electron-phonon, electron-electron Coulomb,
electron-hole Coulomb, and electron-hole exchange (the BAP mechanism)
scatterings are explicitly included. 

We present a phase-diagram--like picture showing the parameter
regime where the DP or BAP mechanism is more important. 
In the case with no impurity and high excitation density, our results
are consistent with  those in  Ref.~\onlinecite{Zhou_PRB_08}: 
i.e., the BAP mechanism is unimportant at low temperature, which is in
stark contrast with the common belief in the
literature.\cite{opt-or,Song,Aronov_83,Fishman,Zerrouati}
However, since we extend the scope of our investigation to higher hole density by
including more hole subbands in the model, 
it is discovered that the BAP mechanism can surpass the DP
mechanism in the regime with high temperature and sufficiently high
hole density. In the cases with low excitation density and/or high
impurity density, the regime where the BAP mechanism surpasses the DP
mechanism becomes larger.
We also show that the temperature regime where
the BAP mechanism is more efficient than the DP mechanism is very
different in the impurity-free and high impurity density cases. 
In the impurity-free case, this regime is around the hole Fermi
temperature for high hole density, regardless of excitation
density. However, in the high impurity density case  
with the identical hole and impurity densities, this regime is
roughly from the electron Fermi temperature to the hole Fermi
temperature. This is because the dominant scatterings in these two
cases are the electron-hole Coulomb scattering and the
electron-impurity scattering, respectively. The different dominant
scatterings lead to the different temperature
dependences of  $\tau_{\rm DP}$, and hence the different 
behaviors of the ratio $\tau_{\rm BAP}/\tau_{\rm DP}$. 
In particular, in the case with high impurity and very low
electron excitation densities, the electron (hole) Fermi temperature
is much lower than (close to) the lowest (highest) temperature of our
investigation. Consequently, the BAP mechanism can dominate the spin
relaxation in the whole temperature regime.
Moreover, we predict that for the impurity-free case, in the regime
where the DP mechanism dominates the spin relaxation, e.g., the cases with high
excitation or low hole density, the total SRT presents a peak 
around the hole Fermi temperature, which is from the nonmonotonic temperature
dependence of the electron-hole Coulomb scattering.

The multi-hole-subband effect on the spin relaxation is also revealed. 
It is shown that the multi-hole-subband effect enhances the relative
importance of the BAP mechanism significantly for high temperature
and/or high hole density. We also predict that at low temperature
the spin relaxation has a very intriguing hole-density dependence thanks to
the contribution from high hole subbands. 
In the impurity-free case, the total SRT shows a double-peak
structure. Both peaks originate from 
the nonmonotonic hole density
dependence of the electron-hole Coulomb scattering.
The only difference is that the first (second) peak
comes from the contribution from the HH$^{(1)}$ (LH$^{(1)}$) subband,
and hence appears around the hole density satisfying $E_{\rm F}^h=k_{\rm B}T$ 
($E_{\rm F}^h-\Delta E_{{\rm LH}^{(1)}}=k_{\rm B}T$).
In the high impurity density case with identical impurity and hole densities, 
there are first a peak and then a valley.
The peak is formed as the DP and BAP
mechanisms compete with each other: the SRT due to the DP (BAP)
mechanism increases (decreases) with $T$ as the DP (BAP) mechanism
dominates at low (high) temperature.
Moreover, since the decrease of the matrix element
$M^-_{-\frac{1}{2},\frac{1}{2}}$ counteracts  the
increase of the other matrix elements of the BAP scattering, 
the hole-density dependence from the
matrix elements is weak. Consequently, when 
the holes in the HH$^{(1)}$ and LH$^{(1)}$ subbands are both in the
degenerate regime, the effect of the increase of the Pauli
blocking is dominant, and $\tau_{\rm BAP}$ increases with $N_h$.
Therefore the valley is formed. However, at high temperature, we show that
the peak from the electron-hole Coulomb scattering disappears and only the
peak from the competition of the BAP and DP mechanisms remains.

Finally, we discuss the relevance to experiments. Since electrons
are minority carriers in $p$-type semiconductors, they have finite
lifetime limited by the electron-hole recombination. From experiments,
the electron lifetime is found to be on the order of several
hundreds of picoseconds,\cite{opt-or,exp} which may be  
shorter than the spin relaxation time in some parameter regime such
as the high impurity density case. However, it has been demonstrated
that spin relaxation time much longer than the lifetime can be
measured via the Hanle or other time-resolved 
measurements.\cite{opt-or,Wagner_93} Hence the calculated long spin relaxation
times can be observed in reality. In short, our
calculation based on realistic parameters provides useful
information in a wide range of experimental conditions, which would
benefit the understanding on spin relaxation.

\begin{acknowledgments}
This work was supported by the National Natural Science Foundation of
China under Grant No.\ 10725417, the National Basic
Research Program of China under Grant No.\ 2006CB922005 and the
Knowledge Innovation Project of Chinese Academy of Sciences.
\end{acknowledgments}

\appendix*

\begin{figure}[htbp]
  \begin{center}
    \includegraphics[height=5.5cm]{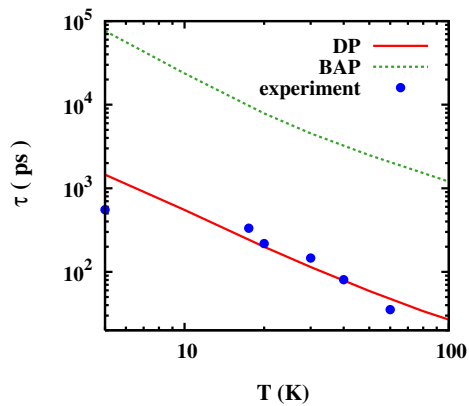}
  \end{center}
  \caption{ (Color online) SRTs $\tau$ from the experimental data in
    Ref.~\onlinecite{exp} and from the DP and BAP mechanisms calculated
    via the KSBE approach in a $p$-type GaAs quantum well.  
   }
  \label{fig_exp}
\end{figure}

\section{Comparison With Experiment}
We compare the calculation via the KSBE approach with the
experimental results in Ref.~\onlinecite{exp}.
In Fig.~\ref{fig_exp}, we plot the temperature dependence of the
SRTs from our computation and from the experiment in $p$-type GaAs
quantum well. Here $a=3$~nm, $N_h=3\times10^{11}$~cm$^{-2}$ and $N_{\rm ex}=2.6\times
10^{9}$~cm$^{-2}$ as indicated in the experiment.\cite{exp} $N_i=0.3N_h$
is obtained by fitting the mobility $\mu\simeq 4000$~cm$^{2}$V$^{-1}$s$^{-1}$ given in
Ref.~\onlinecite{exp}. We take the Dresselhaus spin-orbit coupling parameter as
$\gamma_{\rm D}=18.3$~eV$\cdot$\AA$^{3}$. As the range of $\gamma_{\rm
  D}$ in bulk GaAs calculated and measured via various methods is from
$6.4$ to $25.5$~eV$\cdot$\AA$^{3}$,\cite{SOC1} our value of
$\gamma_{\rm D}$ is reasonable. It is seen that our calculation is in
good agreement with the experimental data. The deviation at $T=5$~K is
likely to be from the fact that the electron temperature $T_e$ is
higher than the lattice temperature $T$ in experiment due to
photo-excitation. This effect becomes significant at low
temperature.\cite{opt-or,Zerrouati} It is also noted that the SRT due
to the BAP mechanism is over one order of magnitude larger
than that from the DP one, i.e., the spin relaxation is totally
governed by the DP mechanism in this case. This is consistent with our
conclusion that the BAP mechanism is unimportant at low impurity and
hole densities.


\begin{thebibliography}{0}
\bibitem{short-range} The contribution from the short-range term is
  negligible.\cite{Jiang_bulk_09}
\bibitem{polarize} Our calculation is not sensitive to the initial
  spin polarization when it is small.
\bibitem{strong_scat} Most cases we discuss in this paper are in the
  strong scattering regime, where $\tau_{\rm DP}$ increases with
  momentum scattering strength and decreases with inhomogeneous
  broadening.
\bibitem{weak_scat} It is seen that in the case with 
  $N_{\rm ex}=10^{11}$ cm$^{-2}$, the variation of the SRT due to the DP
  mechanism becomes very mild when $T<10$ K. This is 
  because this case belongs to the intermediate scattering regime,
  where $\tau_{\rm DP}$ has a weak dependence of the scattering
  strength.
\bibitem{highHH} It is also noted that the contribution
  from the HH$^{(2)}$ subband is always negligible in this case.
\end{thebibliography}
\end{document}